\documentclass[lineno]{JFM-FLM_Au}
\usepackage{xcolor}

\newcommand{\ud}[0]
{
\,\mathrm{d}
}
\newcommand{\q}[2]
{
\frac{\partial #1}{\partial #2}
}
\newcommand{\dd}[2]
{
\frac{\mathrm{d} #1}{\mathrm{d} #2}
}

\DeclareMathOperator\cosec{cosec}

\lefttitle{A. W. Wray}
\righttitle{Resolved multiple-droplet evaporation}

\title{Resolved multiple-droplet evaporation}

\author{Alexander W. Wray}

\affiliation{Department of Mathematics and Statistics, University of Strathclyde, Livingstone Tower, 26 Richmond Street, Glasgow G1 1XH, UK}

\corresau{Alexander W. Wray, alexander.wray@strath.ac.uk}

\begin{document}
\maketitle

\begin{abstract}
A simple, accurate asymptotic formula is derived for the spatially resolved evaporative flux from multiple spherical-cap droplets of arbitrary contact angle.
Kelvin transforms are used to determine the exact response of a spherical-cap droplet to a point source, with extensions to higher multipoles. 
This yields explicit formulae for evaporative shielding and reveals a universal spatial structure in the point-source problem.
Resolving the spatial variation enables the determination of the associated liquid flow and particle transport.
The resulting flux is used to derive an explicit measure of shielding-induced liquid transport.
\end{abstract}



\section{Introduction}\label{sec:intro}

The evaporation of sessile droplets has been extensively studied due to their numerous industrial applications \citep{lohse2020physicochemical}. 
In many situations, the rate-limiting mechanism determining the evaporation is the diffusion of vapour in the atmosphere \citep{hu2002evaporation,wilson2023evaporation}. 
In diffusion-limited evaporation, the evaporative flux is determined by solving a three-dimensional mixed boundary value problem for Laplace's equation.
Few exact solutions are known, and are generally restricted to isolated droplets \citep{popov2005evaporative}.

In many applications, however, droplets occur in arrays, in which neighbouring droplets interact through their vapour fields, an effect studied for at least half a century \citep{labowsky1976effects,schafle1999cooperative,shaikeea2016insight,carrier2016evaporation,khilifi2019study,sadafi2019vapor,edwards2021interferometric,hari2025evaporation}. 
Despite considerable effort, analytical descriptions of this so-called shielding remain limited: existing approaches include theories exploiting the flat-droplet limit \citep{wray2020competitive,wray2024high} and approximate interaction models \citep{hatte2019universal,tonini2022analytical,tonini2024modeling,argatov2026modified}. 
Some studies focus on integral evaporation rates \citep{masoud2021evaporation,azzam2024modeling}, but in so doing do not resolve the non-axisymmetric local flux needed to model liquid transport and deposition \citep{deegan1997capillary,pradhan2015deposition,lee2023vapor,wray2021contact}. 
Studies of liquid transport in droplet arrays have therefore generally used numerical solutions or approximate evaporation laws \citep{debnath2026translation, dekker2026internal}.
Here we give an asymptotic derivation for the spatially varying flux from arrays of spherical-cap droplets of arbitrary contact angle. 
%


\section{Problem formulation}\label{sec:probForm}

Consider $N$ spherical-cap droplets evaporating in proximity, with contact radii $a_i$ and contact angles $0<\theta_i<\pi$, whose footprint centres are located at $\mathbf{X}_{i}=(x_i,y_i,0)$ on the substrate $z=0$ for $1\leq i\leq N$.
Let $\Gamma_i$ denote the free surface of droplet $i$ and $\Gamma_s$ the unwetted substrate.
In the standard nondimensionalisation of \citet{wilson2023evaporation}, the evaporative flux from the surface of droplet $i$ is given by
\begin{equation}
    J_i(\mathbf{x})=-\mathbf{n}\cdot\nabla c({\mathbf{x}}), \qquad \mathbf{x}\in\Gamma_i,
\end{equation}
where $\mathbf{x}=(x,y,z)$, $\mathbf{n}$ is the outward unit normal from the liquid, and $c$ is the dimensionless concentration of vapour in the atmosphere, which satisfies
\begin{equation}
    \nabla^2c=0,
\end{equation}
subject to
\begin{equation}
    c=1 \quad \text{on $\Gamma_i$,} \quad 
    c_z=0 \quad \text{on $\Gamma_s$,} \quad 
    c\to 0 \quad \text{as $|\mathbf{x}|\to\infty$.}
\end{equation}
The total evaporation rates, determined in Section \ref{sec:multipleSources} by integrating the local fluxes, are denoted by
\begin{equation}
    F_i=\iint_{\Gamma_i} J_i(\mathbf{x})\,\ud S.
\end{equation}
As $c_z=0$ on $\Gamma_s$, it is equivalent to solve the problem in which the substrate is removed and each droplet is complemented by its reflection across $z=0$ \citep{popov2005evaporative}.
The distance from a point $\mathbf{x}$ to the centre of the footprint of droplet $j$ is denoted by
\begin{equation}
    \ell_j(\mathbf{x})=|\mathbf{x}-\mathbf{X}_j|.
\end{equation}
Let $c_j^\mathrm{iso}$ denote the concentration field of droplet $j$ in isolation, and $F_j^{\mathrm{iso}}$ its total evaporation rate. 
The \emph{intrinsic} contribution $c_j^\mathrm{intr}=(F_j/F_j^{\mathrm{iso}})c_j^\mathrm{iso}$ has the far-field multipole expansion
\begin{equation}
    c_j^\mathrm{intr}(\mathbf{x})
    =q_j\left[ 
        \frac{1}{\ell_j}+\beta_j\left( \frac{\ell_j^2-3z^2}{\ell_j^5} \right) +O\left( \ell_j^{-5}\right)
    \right],\label{eq:multipoleExp}
\end{equation}
where 
\begin{align}
    q_j&=\frac{F_j}{2\pi}, \qquad \beta_j=a_j^2\frac{A(\theta_j)-4B(\theta_j)}{4A(\theta_j)},\\
    \left( A(\theta),B(\theta) \right)&=\int_0^\infty \frac{(1,\tau^2)}{1+\cosh\left[ (2\pi-\theta)\tau \right]/\cosh(\theta\tau)}\ud\tau, 
\end{align}
\citep{kek1992dropwise,masoud2021evaporation}.
The terms in \eqref{eq:multipoleExp} are the intrinsic monopole and quadrupole, as axisymmetry and reflection symmetry eliminate odd multipoles.

The neighbouring droplets also generate an \emph{induced} contribution which is not present in \eqref{eq:multipoleExp}. 
In particular, the incident concentration field from other droplets induces a horizontal dipole whose strength is proportional to the concentration gradient. 
If the droplets are well separated and have dimensions of order unity at most then, defining
\begin{equation}
    \epsilon=\max_{i\neq j} \frac{1}{b_{ij}}\ll1, \quad \text{where} \quad  b_{ij}=|\mathbf{X}_{i}-\mathbf{X}_j|,
\end{equation}
the strength of this induced dipole is at most $O(\epsilon^{2})$. 
Its contribution at the surface of another droplet, where $\ell_{j}\gtrsim \epsilon^{-1}$, is then at most $O(\epsilon^2\ell_j^{-2})\leq O(\epsilon^4)$, 
so that the induced dipole contribution is generally the next largest contribution after the intrinsic quadrupole.
We begin in Section \ref{sec:exactSol} by assuming that other droplets can be treated as point sources (i.e. their leading-order contribution in \eqref{eq:multipoleExp}), and determining the complete induced response. 
This point-source assumption is common in the evaporating-droplet literature \citep{wray2024high,azzam2024modeling}, and, as shown below, is equivalent to the approach used by \citet{wray2020competitive} for flat droplets. 
Higher-order corrections are discussed in Section \ref{sec:higher}.


\section{Exact solutions}\label{sec:exactSol}

\subsection{Single source}\label{sec:singleSource}

\begin{figure}
\centering{\includegraphics[width=0.6\textwidth]{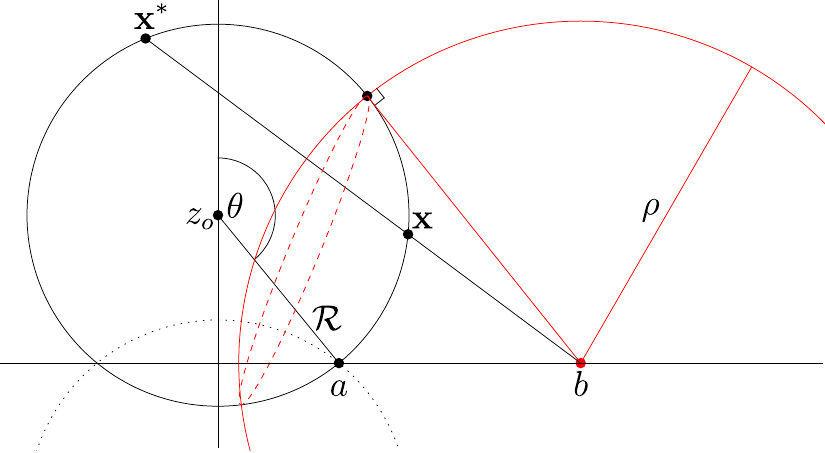}}
\caption{
Geometry of the point-source problem in Section \ref{sec:singleSource}. 
The points $\mathbf{x}$ and $\mathbf{x}^\ast$ are inverses of one another, the dashed red curve denotes the tangent ring, and the dotted black curve denotes the image sphere.
}
\label{fig:schematicSetup} 
\end{figure}

We first consider a pair of droplets, with $i$ denoting the target droplet and $j$ the source droplet, replace the latter by a point source, and work relative to an origin located at $\mathbf{X}_i$.
For brevity, we suppress the indices, with $q=q_j$ and $\beta=\beta_j$ referring to the source, and other droplet quantities to the target. 
We denote the position of the source by $\mathbf{b}=b_x\mathbf{e}_x+b_y\mathbf{e}_y$, and rotate the coordinate system so that the source is located at $\mathbf{b}=b\mathbf{e}_x$, with $b=b_{ij}>a$.
The sphere containing the droplet surface has radius and centre height
\begin{equation}
    \mathcal{R}=a\cosec \theta, \quad z_o=-a\cot \theta. 
\end{equation}
The geometry is shown in Figure \ref{fig:schematicSetup}. 
Defining
\begin{equation}
    \ell(\mathbf{x})=|\mathbf{x}-\mathbf{b}|,
\end{equation}
the incident concentration field generated by the point source is
\begin{equation}
    c^\mathrm{src}(\mathbf{x})=\frac{q}{\ell(\mathbf{x})},\label{eq:pointChargeTarget}
\end{equation}
and the target-droplet response must be chosen so as to ensure the total concentration remains unity on the droplet surface.

We seek a Kelvin inversion of radius $\rho$ centred on $\mathbf{b}$ that leaves both the target and image spheres invariant,
\begin{equation}
    \mathbf{x}^\ast(\mathbf{x})=\mathbf{b}+\frac{\rho^2}{\ell^2}(\mathbf{x}-\mathbf{b}). \label{eq:inversion}
\end{equation}
This corresponds to inversion in the sphere $S(\mathbf{b},\rho)$, under which $S(z_o\mathbf{e}_z,\mathcal{R})$ is mapped to
\begin{equation}
    S\left(\mathbf{b}+\frac{\rho^2(z_o\mathbf{e}_z-\mathbf{b})}{b^2-a^2},\frac{\mathcal{R}\rho^2}{b^2-a^2} \right)
\end{equation}
\citep{amaral2017kelvin}. 
Hence, choosing $\rho^2=b^2-a^2$
leaves both the centre and the radius of the sphere unchanged. 
The same applies to the image sphere with $z_o$ replaced with $-z_o$, so the entire geometry is invariant.

The isolated concentration field 
satisfies $c^{\mathrm{iso}}|_\Gamma=1$, with evaporative flux
\begin{equation}
    J^{\mathrm{iso}}(\mathbf{x})=-\q{c^{\mathrm{iso}}}{n}.
\end{equation}
By axisymmetry, on $\Gamma$ we may equivalently write $J^{\mathrm{iso}}(\mathbf{x})=J^{\mathrm{iso}}(z)$.
The corresponding Kelvin transform of $c^{\mathrm{iso}}$, denoted $c^{\mathrm{iso},\ast}$, satisfies
\begin{equation}
    c^{\mathrm{iso},\ast}(\mathbf{x})=\frac{1}{\ell}c^{\mathrm{iso}}\left(\mathbf{x}^\ast\right),
\end{equation}
and, as $c^{\mathrm{iso}}$ is harmonic, so too is $c^{\mathrm{iso},\ast}$. 
For $\mathbf{x}\in \Gamma$, 
\begin{equation}
    c^{\mathrm{iso},\ast}(\mathbf{x})=\frac{1}{\ell}.
\end{equation}
Hence $c^{\mathrm{iso},\ast}$ is a harmonic function with boundary data identical to that induced by a unit point source at $\mathbf{b}$ \eqref{eq:pointChargeTarget}.

Therefore, the concentration field in the presence of a point source of strength $q$ is
\begin{align}
c^{\mathrm{ps}}(\mathbf{x})&=c^{\mathrm{iso}}(\mathbf{x})+\frac{q}{\ell}\left[1-c^{\mathrm{iso}}\left(b\mathbf{e}_x+\frac{\rho^2(\mathbf{x}-b\mathbf{e}_x)}{\ell^2}\right)\right],
\end{align}
and the corresponding evaporative flux is
\begin{equation}
    J^{\mathrm{ps}}=J^{\mathrm{iso}}(\mathbf{x})-q\frac{\rho^2}{\ell^3}J^{\mathrm{iso}}(\mathbf{x}^\ast).\label{eq:exactSol}
\end{equation}
Equation \eqref{eq:exactSol} is the key analytical result of this paper: given $J^\mathrm{iso}$, the exact point-source response requires only one additional evaluation at the inversion point $\mathbf{x}^\ast$. 
This solution is exact for arbitrary contact angle $0<\theta<\pi$ and point source distance $b>a$,
although the validity of truncating \eqref{eq:multipoleExp} decreases as $b$ decreases.
$J^\mathrm{iso}$ is known in elementary closed form for flat droplets and contact angles of the form $\pi(1-1/(2n)), n=1,2,\ldots$ \citep{lebedev2023evaporation}; for arbitrary contact angle, it can be evaluated using the expressions of \citet{popov2005evaporative}.
The evaporative flux \eqref{eq:exactSol} can also be written as
\begin{equation}
    J^{\mathrm{ps}}=J^{\mathrm{iso}}(\mathbf{x})\left[1-S(\mathbf{x}) \right], \quad S(\mathbf{x})=q\frac{\rho^2}{\ell^3}\frac{J^{\mathrm{iso}}(\mathbf{x}^\ast)}{J^{\mathrm{iso}}(\mathbf{x})},\label{eq:shieldingFactor}
\end{equation}
so that $S(\mathbf{x})$ is the shielding factor.
\eqref{eq:inversion} gives $z^\ast=(\rho^2/\ell^2)z$, so that
\begin{equation}
    S(\mathbf{x})=\frac{q}{\rho}\frac{\Phi(z^\ast)}{\Phi(z)},  \quad \Phi(z)=z^{3/2}J^{\mathrm{iso}}(z), \label{eq:shieldPForm}
\end{equation}
so that the shielding depends only on the vertical position of the point and its inverse. 

\subsection{Shielding}\label{sec:shielding}

The spatial dependence of the shielding behaviour can now be characterised.
The self-inverse points, which satisfy $\mathbf{x}=\mathbf{x}^\ast$, form a ring on the sphere surface.
We term this the tangent ring, although only some portion may intersect the spherical cap.
On this ring,
\begin{equation}
    \ell=\rho, \quad S=\frac{q}{\rho}, \quad J^{\mathrm{ps}}(\mathbf{x})=J^{\mathrm{iso}}(z)\left[ 1-\frac{q}{\rho} \right].
\end{equation}

The large-$b$ expansion of \eqref{eq:exactSol} is
\begin{equation}
    J^{\mathrm{ps}}=J^{\mathrm{iso}}\left(1-\frac{q}{b} \right)-2\frac{qx}{b^2\sqrt{z}}\dd{\Phi}{z}+O(b^{-3}).\label{eq:ffExp}
\end{equation}
The leading relative correction $q/b$ is spatially uniform, while the $O(b^{-2})$ term represents the response to the weak horizontal concentration gradient generated by the distant point source. 
At a given height this term is proportional to $x$, and so we need only determine the sign of its coefficient. 
Since the gradient suppresses evaporation more strongly on the source-facing side, \eqref{eq:ffExp} implies $\mathrm{d}\Phi/\mathrm{d}z>0$ (as may be confirmed using the maximum principle and Hopf's lemma). 
Then, differentiating \eqref{eq:shieldPForm} at fixed $z$ yields
\begin{equation}
    \left.\q{S}{\ell}\right|_z=-\frac{2q\rho z}{\ell^3}\frac{\Phi'(z^\ast)}{\Phi(z)}, \label{eq:shieldingVsPSDistance}
\end{equation}
so that $S$ decreases monotonically with distance from the point source at a fixed height $z$.
The tangent ring therefore divides the droplet surface into regions nearer to and further from the source, satisfying $S>q/\rho$ and $S<q/\rho$, respectively.
Inverse points fall on opposite sides of this ring and, because $S(\mathbf{x})S(\mathbf{x}^\ast)=q^2/\rho^2$,
their shielding factors satisfy
\begin{equation}
    \log \frac{\rho S(\mathbf{x}^\ast)}{q}=-\log \frac{\rho S(\mathbf{x})}{q}.
\end{equation}
This asymmetric evaporation drives liquid flow and resulting particle transport (see Section \ref{sec:transport}).

Close to the contact line,
\begin{equation}
    J^{\mathrm{iso}}\sim K z^{-\lambda}, \quad \lambda=\frac{\pi-2\theta}{2\pi-2\theta},
\end{equation}
for some $K>0$. 
Hence, from \eqref{eq:shieldingFactor},
\begin{equation}
    S(\mathbf{x})\sim q\frac{\rho^{2(1-\lambda)}}{\ell^{3-2\lambda}},
\end{equation}
so that, in particular, the contact-line flux exponent is unchanged whenever the limiting value of the shielding is not unity.
This relative correction is explicit for arbitrary $\theta$, even when $J^\mathrm{iso}$ is not. 
In particular, the ratio of the shielding at the contact line at the points nearest to and furthest from the point source is
\begin{equation}
    \frac{(b+a)^{3-2\lambda}}{(b-a)^{3-2\lambda}}=\left( \frac{b+a}{b-a} \right)^{1+\frac{\pi}{\pi-\theta}}.
\end{equation}

For a flat droplet
\begin{equation}
    J^{\mathrm{ps}}(\mathbf{x})=J^{\mathrm{iso}}(\mathbf{x})\left[ 1-q\frac{\sqrt{b^2-a^2}}{\ell^2} \right],
\end{equation}
which coincides exactly with the result of \citet{wray2020competitive}. 
At the special dewetting angles noted in Section \ref{sec:singleSource}, the problem admits classical image solutions
\citep{lindell2003electrostatic}.

\subsection{Higher-order corrections}\label{sec:higher}

The correction due to the next contribution, namely the intrinsic quadrupole, can be determined by replacing \eqref{eq:pointChargeTarget} with
\begin{equation}
    c^{\mathrm{src}}=q\left[ \frac{1}{\ell}+\beta\frac{\ell^2-3z^2}{\ell^5} \right]=q\left[
        \frac{1}{\ell}+\beta \nabla_\mathbf{b}^2\left( \frac{1}{\ell} \right)
    \right],
\end{equation}
where $\nabla_\mathbf{b}^2=\partial_{b_x}^2+\partial_{b_y}^2$. 
Since $\mathbf{b}$ affects only the source position, not the target droplet or its boundary conditions, $\nabla_\mathbf{b}^2$ commutes with solving for the point source response, and hence
\begin{equation}
    J^\mathrm{qp}=J^{\mathrm{iso}}-q\left( 1+\beta \nabla_\mathbf{b}^2 \right)\left( \frac{\rho^2}{\ell^3}J^\mathrm{iso}(\mathbf{x}^\ast) \right).\label{eq:secondOrderSol}
\end{equation}
This expression is more complicated than \eqref{eq:exactSol} and does not share its convenient simple properties. 
However, $A(\theta_j)$ and $B(\theta_j)$ are evaluated when determining the $F_i$, and so the only additional step in computing the local flux is the numerical evaluation of the $\nabla_\mathbf{b}^2$ term.

\subsection{Multiple sources}\label{sec:multipleSources}
Define
\begin{equation}
    \mathbf{b}_{ij}=\mathbf{X}_j-\mathbf{X}_{i},
    \qquad
    b_{ij}=|\mathbf{b}_{ij}|,
    \qquad
    \rho_{ij}^2=b_{ij}^2-a_i^2,
    \qquad
    \mathbf{x}_{ij}^{*}
    =
    \mathbf{X}_j+
    \frac{\rho_{ij}^2}{\ell_j^2(\mathbf{x})}
    (\mathbf{x}-\mathbf{X}_j).\label{eq:multiInv}
\end{equation}
The shielding of droplet $i$ by source $j$ is
\begin{equation}
    S_{ij}(\mathbf{x})
    =q_j\frac{\rho_{ij}^2}{\ell_j(\mathbf{x})^3}
    \frac{J_i^{\mathrm{iso}}(\mathbf{x}_{ij}^\ast)}
         {J_i^{\mathrm{iso}}(\mathbf{x})},
\end{equation}
so that
\begin{equation}
    J_i(\mathbf{x})
    =J_i^{\mathrm{iso}}(\mathbf{x})[1-S_i(\mathbf{x})],
    \qquad
    S_i(\mathbf{x})=\sum_{j\neq i}S_{ij}(\mathbf{x}).\label{eq:multidropFormula}
\end{equation}
Under the inversion \eqref{eq:multiInv}, the surface area element transforms as $\ell_j^4\ud S^\ast=\rho_{ij}^4 \ud S$. 
The standard single-layer representation then gives
\begin{equation}
    \iint_{\Gamma_i} \frac{\rho_{ij}^2}{\ell_j^3}J_i^\mathrm{iso}(\mathbf{x}_{ij}^\ast)\ud S=\iint_{\Gamma_i} \frac{J_i^\mathrm{iso}(\mathbf{x})}{\ell_j}\ud S=2\pi C_i^\mathrm{iso}(b_{ij}),
\end{equation}
where $C_i^{\mathrm{iso}}(r)=c_i^{\mathrm{iso}}(\mathbf X_i+r\mathbf e_x)$ is the isolated concentration field on the substrate. 
Therefore, integrating \eqref{eq:multidropFormula} over the surface of droplet $i$ yields
\begin{equation}
    F_i=F_i^{\mathrm{iso}}-\sum_{j\neq i}F_j C_i^{\mathrm{iso}}(b_{ij}).\label{eq:FRelation}
\end{equation}
\eqref{eq:FRelation} is a linear system for the $F_i$, of the form derived by \citet{fabrikant1987diffusion} and, up to the choice of evaluation point, \citet{masoud2021evaporation}.
At higher order, \eqref{eq:secondOrderSol} yields
\begin{align}
J_i^{\mathrm{qp}}(\mathbf{x})
    &=J_i^{\mathrm{iso}}(\mathbf{x})
    -\sum_{j\neq i}q_j
    \left(1+\beta_j\nabla_{\mathbf{b}_{ij}}^2\right)
    \left[
        \frac{\rho_{ij}^2}{\ell_j(\mathbf{x})^3}
        J_i^{\mathrm{iso}}(\mathbf{x}_{ij}^\ast)
    \right],
    \label{eq:multipleQp}\\
    F_i&=F_i^{\mathrm{iso}}-\sum_{j\neq i}F_j\left[ C_i^{\mathrm{iso}}(b_{ij})+\beta_j\left( {C_i^{\mathrm{iso}}}''(b_{ij})+\frac{{C_i^{\mathrm{iso}}}'(b_{ij})}{b_{ij}} \right) \right].
\end{align}
$C_i^{\mathrm{iso}}(b_{ij})$ and its derivatives may be evaluated using \eqref{eq:multipoleExp}, retaining terms up to $b_{ij}^{-3}$.


\section{Validation and results}\label{sec:valid}

We validate and investigate the leading-order \eqref{eq:exactSol} and higher-order \eqref{eq:secondOrderSol} solutions. 
Reference solutions were computed in Mathematica using the finite-element method (FEM), with quadratic elements in a cuboidal domain. 
Surface fluxes were obtained using second-order one-sided normal differences. 
Convergence was checked by enlarging the domain and refining the mesh and flux calculations. 
In all cases, each droplet has maximal lateral extent unity, corresponding to $a_i=1$ for $\theta_i\leq\pi/2$ and $\mathcal{R}_i=1$ for $\theta_i>\pi/2$.

\begin{figure}
\centering 
    \begin{tabular}{cc}
\includegraphics[width=0.45\textwidth]{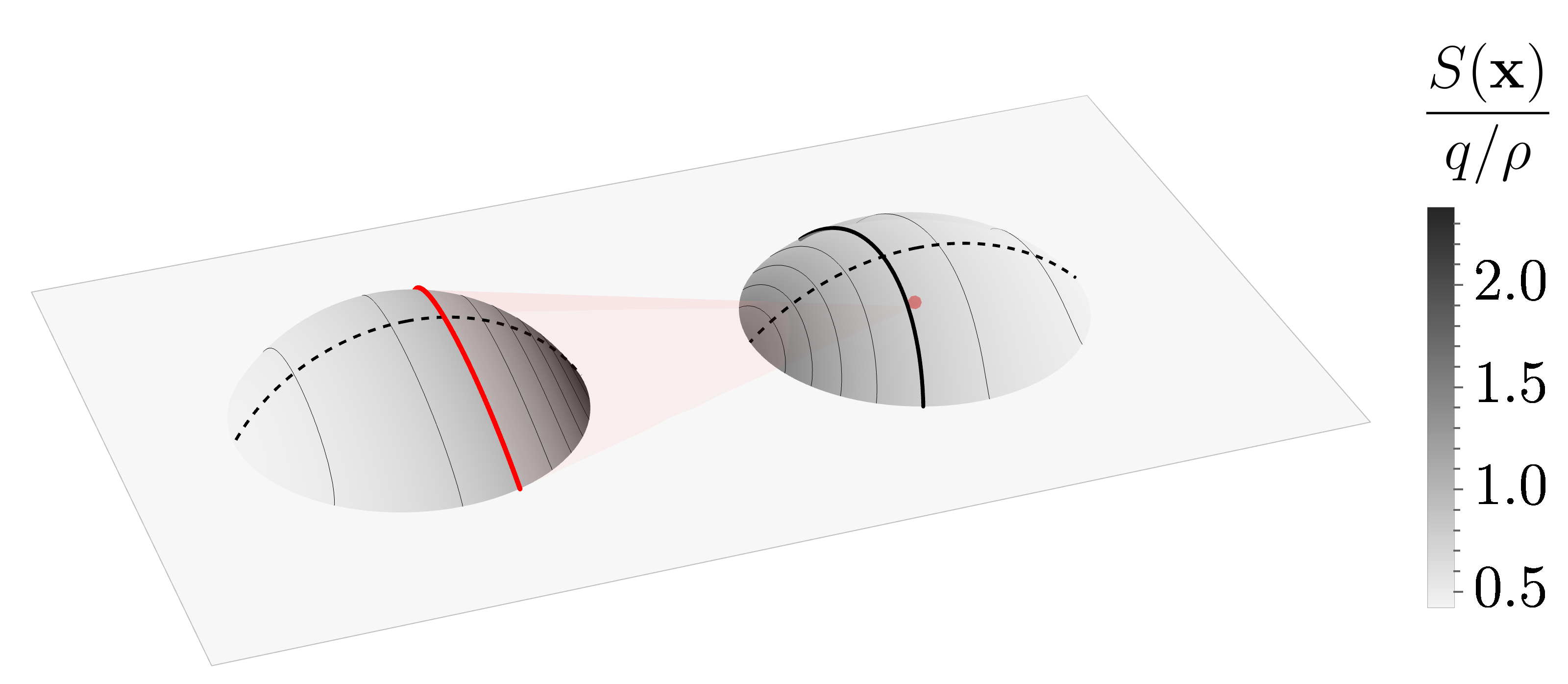} &
\includegraphics[width=0.45\textwidth]{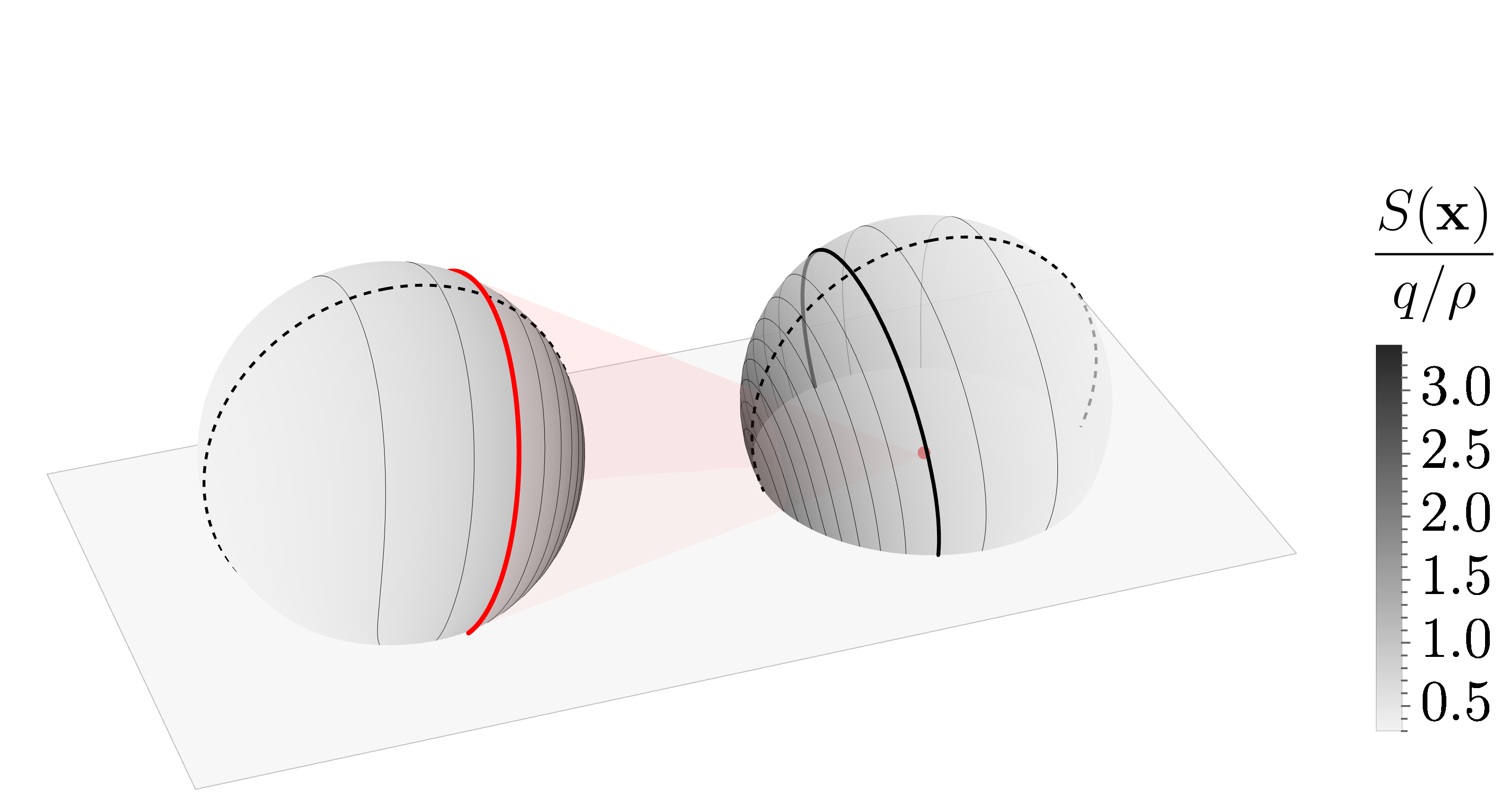} \\
(a) & (b) \\
\includegraphics[width=0.45\textwidth]{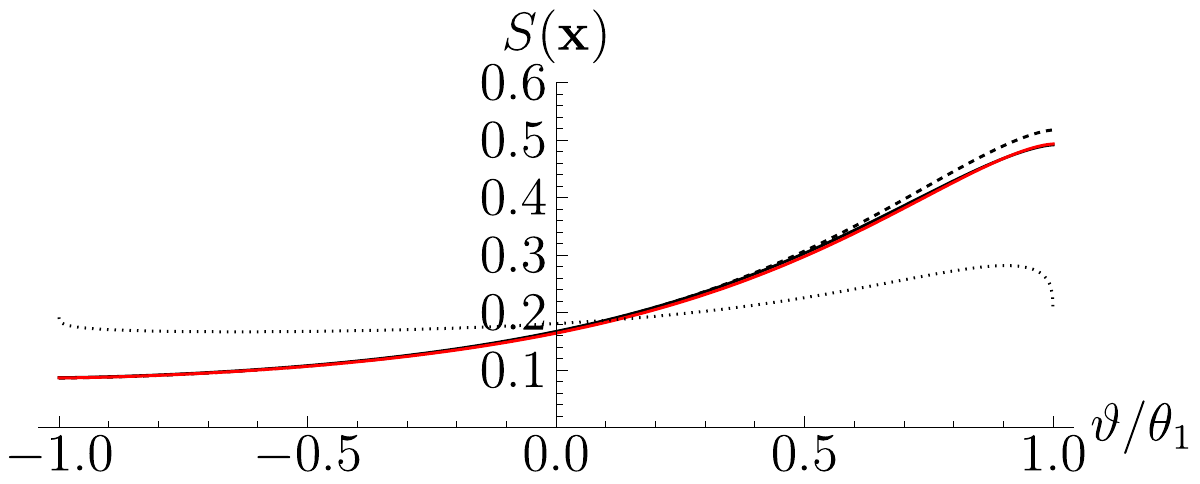} &
\includegraphics[width=0.45\textwidth]{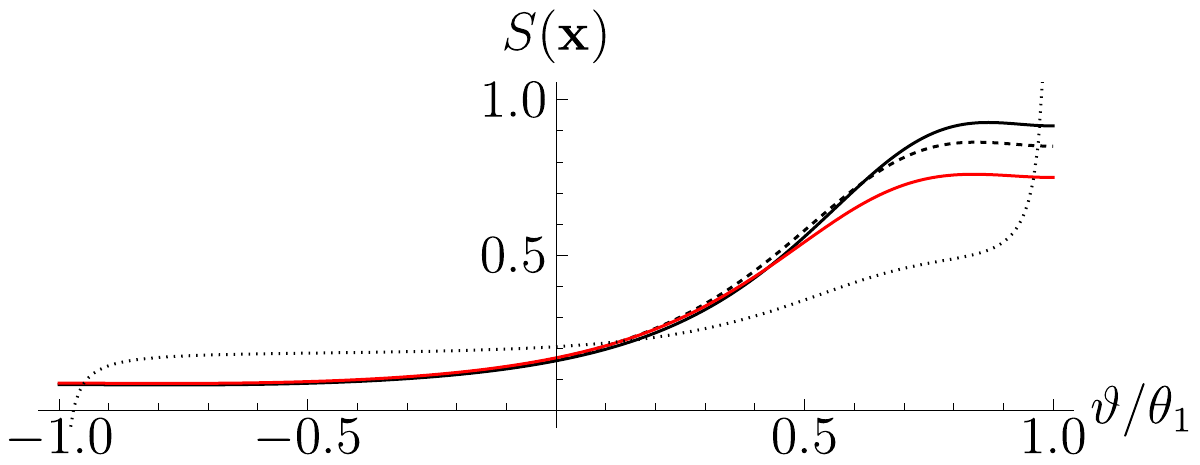} \\
(c) & (d) \\
\includegraphics[width=0.45\textwidth]{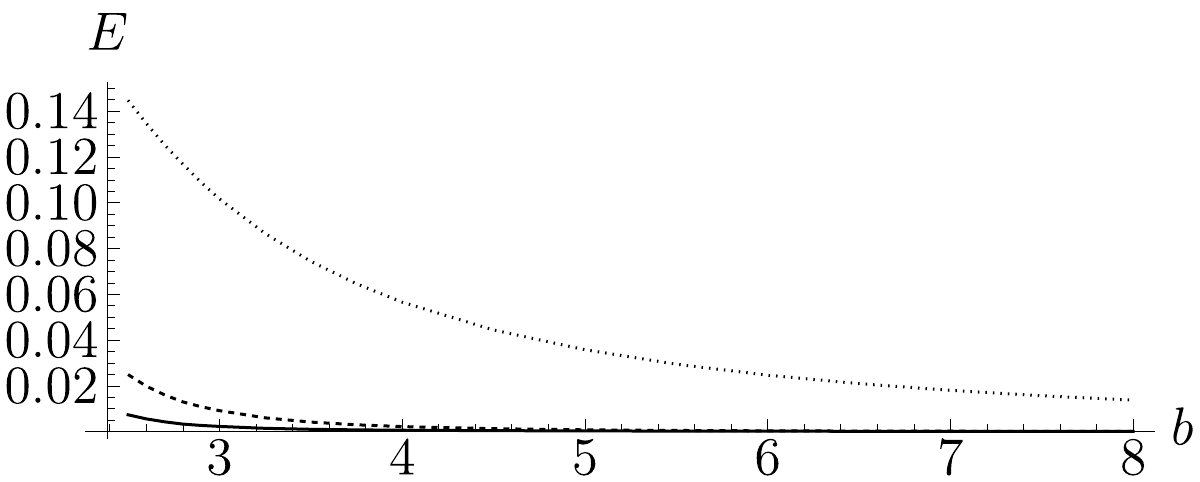} &
\includegraphics[width=0.45\textwidth]{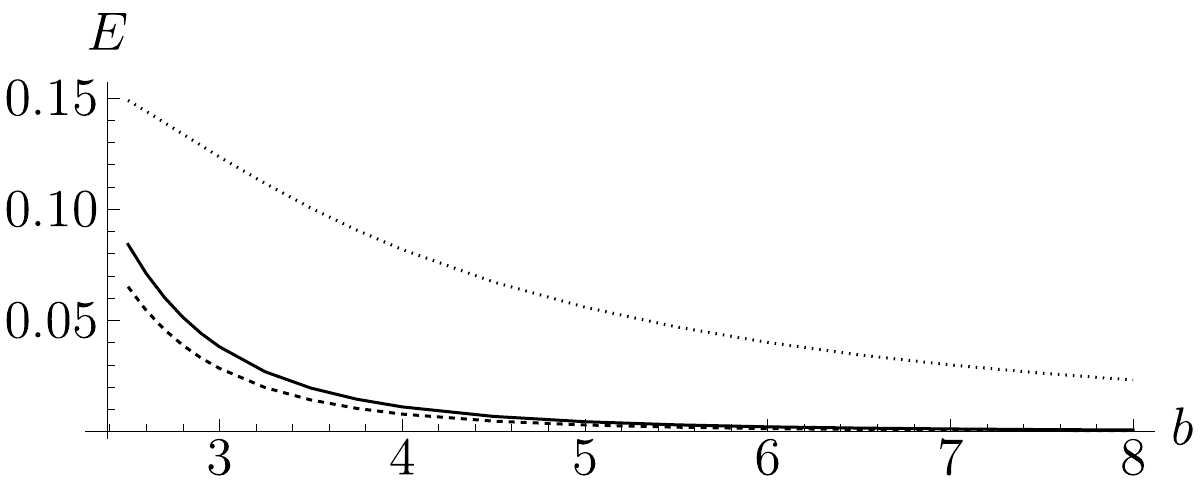} \\
(e) & (f) 
\end{tabular}
\caption{
        Shielding of pairs of droplets with 
        (a,c,e) $(\theta_1,\theta_2)=(\pi/3,\pi/4)$; 
        (b,d,f) $(\theta_1,\theta_2)=(3\pi/4,5\pi/8)$.
        In (a-d), $b=3$.
        (a,b) Shielding of both droplets according to the leading-order solution \eqref{eq:exactSol}, showing the tangent ring (red), centreline (dashed), and contours at intervals of $0.25$ in $S/(q/\rho)$;
        (c,d) shielding along the centreline of droplet 1, where $\vartheta$ is the polar angle;
        (e,f) relative error \eqref{eq:JErr} against $b$.
        In (c-f), 
        solid black lines correspond to the leading-order solution $J^\mathrm{ps}$ \eqref{eq:exactSol}, 
        dashed black lines the higher-order solution $J^\mathrm{qp}$ \eqref{eq:secondOrderSol},
        and dotted black lines the model of \citet{tonini2024modeling}.
        Red lines in (c,d) show the FEM results.
    }\label{fig:twoDroplets}
\end{figure}

Figure \ref{fig:twoDroplets} compares shielding for pairs of wetting and dewetting droplets.
Figures \ref{fig:twoDroplets}(a,b) show $S(\mathbf{x})$ normalised by its value $q/\rho$ on the tangent ring.
The red cone is tangent to the droplet along this ring, so that $S>q/\rho$ on the side nearer the source and $S<q/\rho$ on the far side.
Figures \ref{fig:twoDroplets}(c,d) show the shielding along the centreline of droplet $1$, while (e,f) show the relative integral square surface error
\begin{equation}
E=\left[\frac{\iint_{\Gamma_1} |J^m-J_\mathrm{FEM}|^2\ud S}
    {\iint_{\Gamma_1} |J_\mathrm{FEM}|^2\ud S}\right]^{1/2},
    \label{eq:JErr}
\end{equation}
where $J_\mathrm{FEM}$ and $J^m$ are the finite-element and respective model fluxes.
Few models are available for thick evaporating droplets, so we also compare with the superposition model of \citet{tonini2024modeling}, applied here to droplets with unequal contact angles.
For the two pairs shown, the errors in \eqref{eq:exactSol} and \eqref{eq:secondOrderSol} decrease with increasing $b$ and remain below $9\%$ over the range shown.
The errors are largest for the closest dewetting cases, which provide the most stringent test of the point-source approximation.
For wetting droplets at small $b$, the leading-order solution \eqref{eq:exactSol} can give better agreement than the higher-order one \eqref{eq:secondOrderSol}; partial cancellation between the intrinsic-quadrupole and omitted induced-dipole corrections may explain this behaviour. 
For still smaller values of $b$, the point-source approximation can eventually break down, with predicted shielding locally exceeding unity.

\begin{figure}
\centering 
    \begin{tabular}{ccc}
\includegraphics[width=0.32\textwidth]{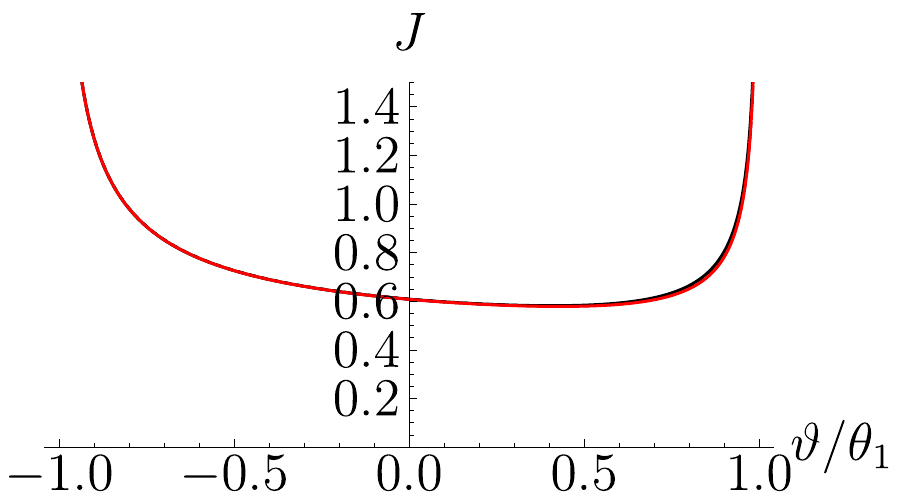} &
\includegraphics[width=0.32\textwidth]{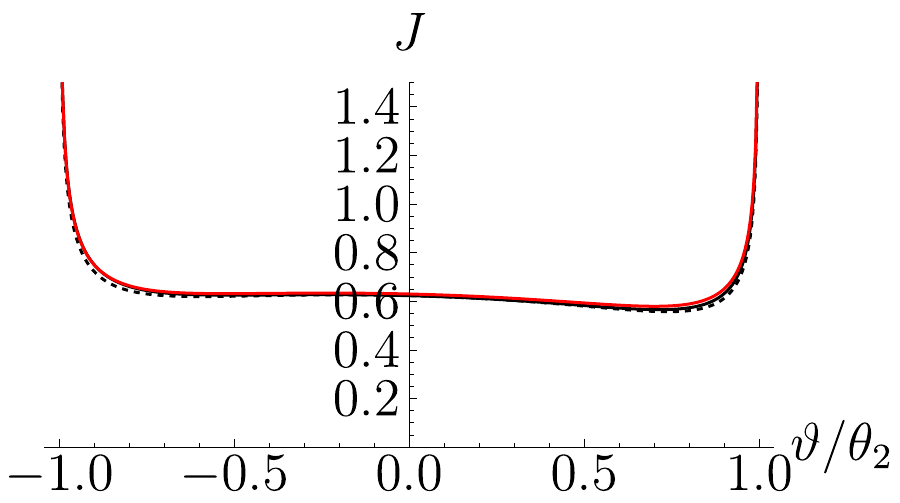} &
\includegraphics[width=0.32\textwidth]{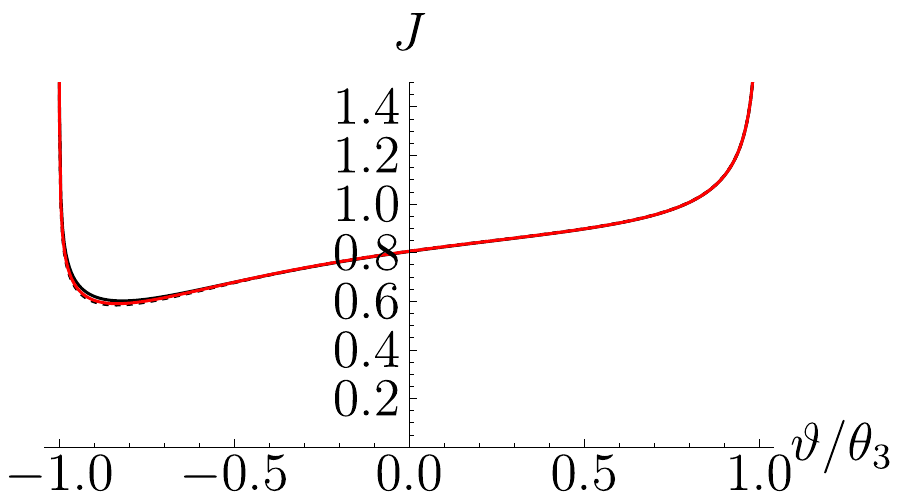} \\
(a) & (b) & (c) \\
\includegraphics[width=0.32\textwidth]{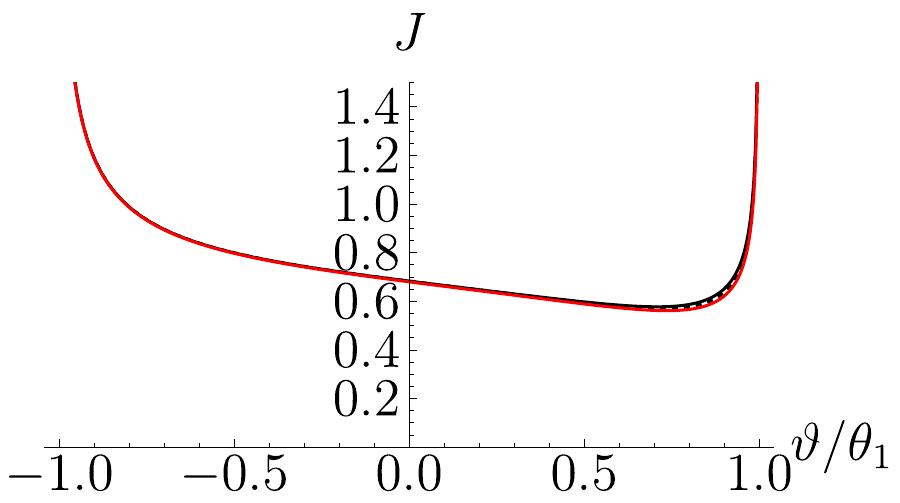} &
\includegraphics[width=0.32\textwidth]{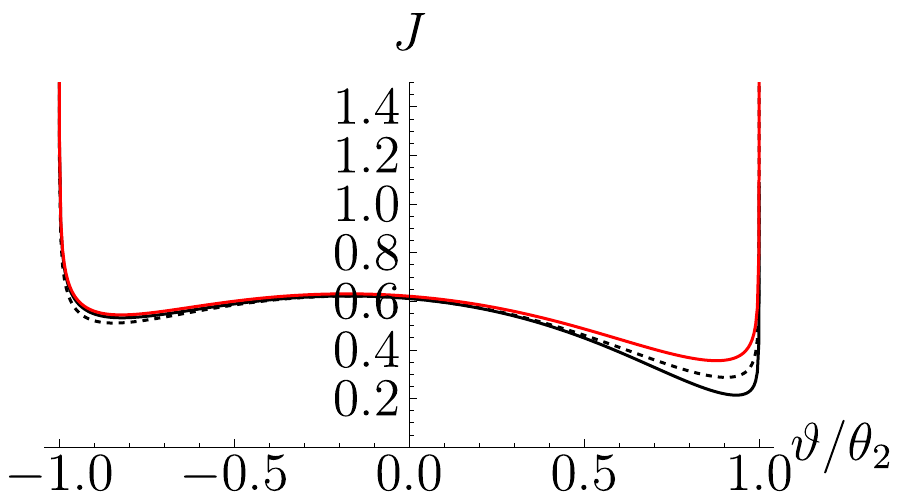} &
\includegraphics[width=0.32\textwidth]{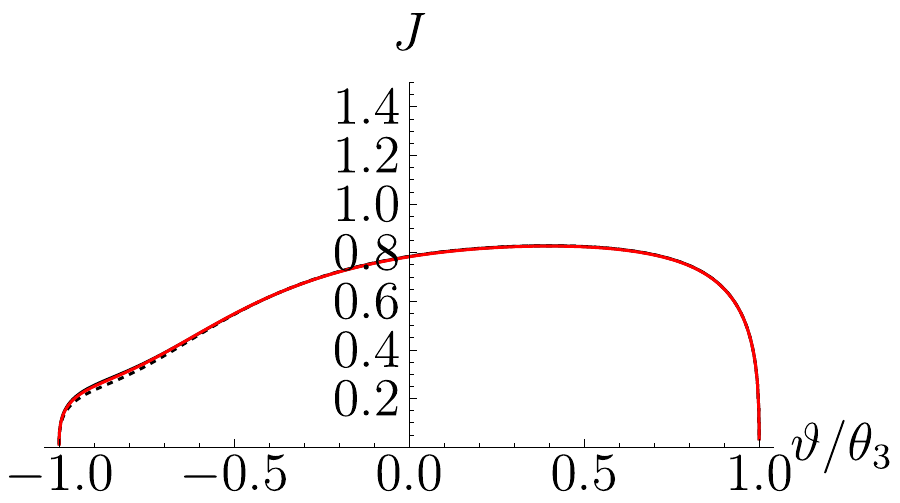} \\
(d) & (e) & (f) 
    \end{tabular}
    \caption{
        Evaporative flux along $y=0$ for three droplets located at $(-3,0)$, $(0,0)$ and $(3,0)$. 
        Top row: (a) $\theta_1=\pi/8$, (b) $\theta_2=\pi/4$, (c) $\theta_3=3\pi/8$; 
        bottom row: (d) $\theta_1=\pi/4$, (e) $\theta_2=3\pi/8$, (f) $\theta_3=5\pi/8$. 
        Solid black lines show the leading-order solution \eqref{eq:multidropFormula}, dashed black lines the higher-order solution \eqref{eq:multipleQp}, and red lines the FEM results.
    }\label{fig:verifThreeDroplets}
\end{figure}

Figure \ref{fig:verifThreeDroplets} compares the evaporative flux along $y=0$ for three droplets, showing generally good agreement.
The largest discrepancy occurs on the side of droplet 2 facing the dewetting droplet 3, shown in Figure \ref{fig:verifThreeDroplets}(e). 
Incorporating the intrinsic quadrupole in \eqref{eq:multipleQp} improves agreement there. 

\begin{figure}
\centering 
    \begin{tabular}{cc}
\includegraphics[width=0.41\textwidth]{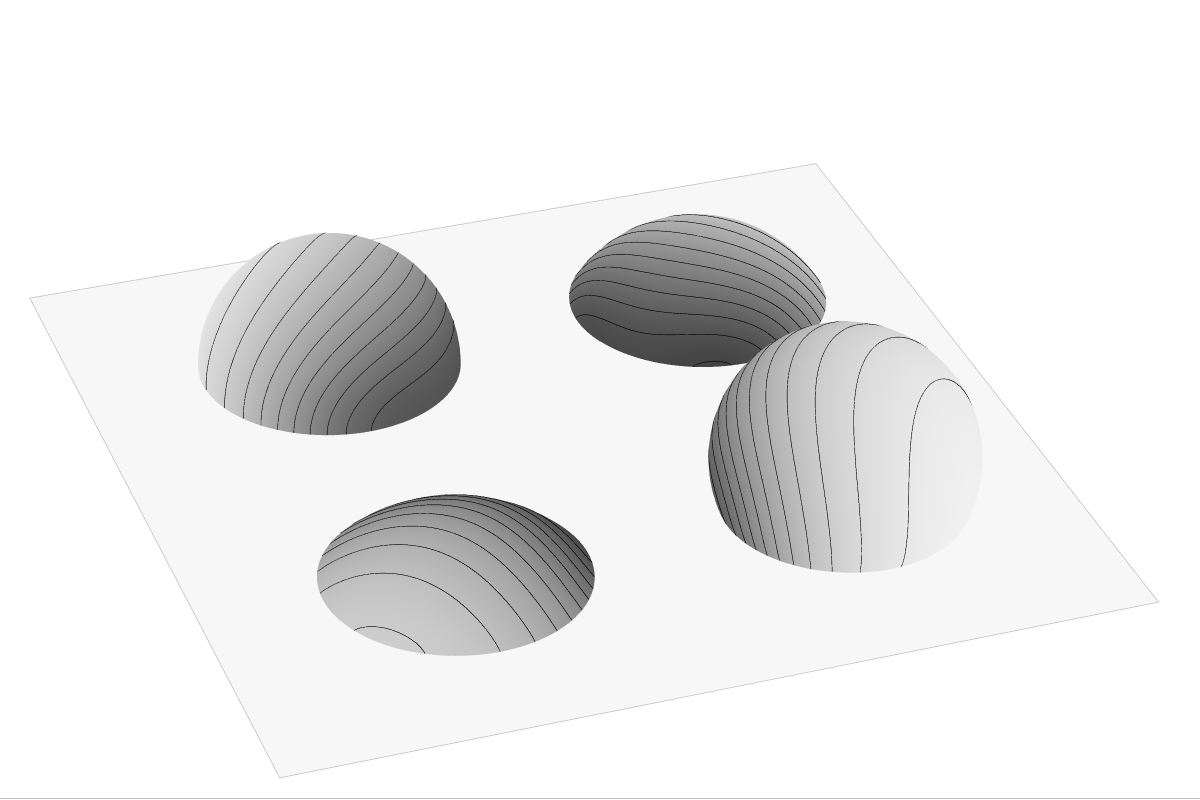} &
\includegraphics[width=0.41\textwidth]{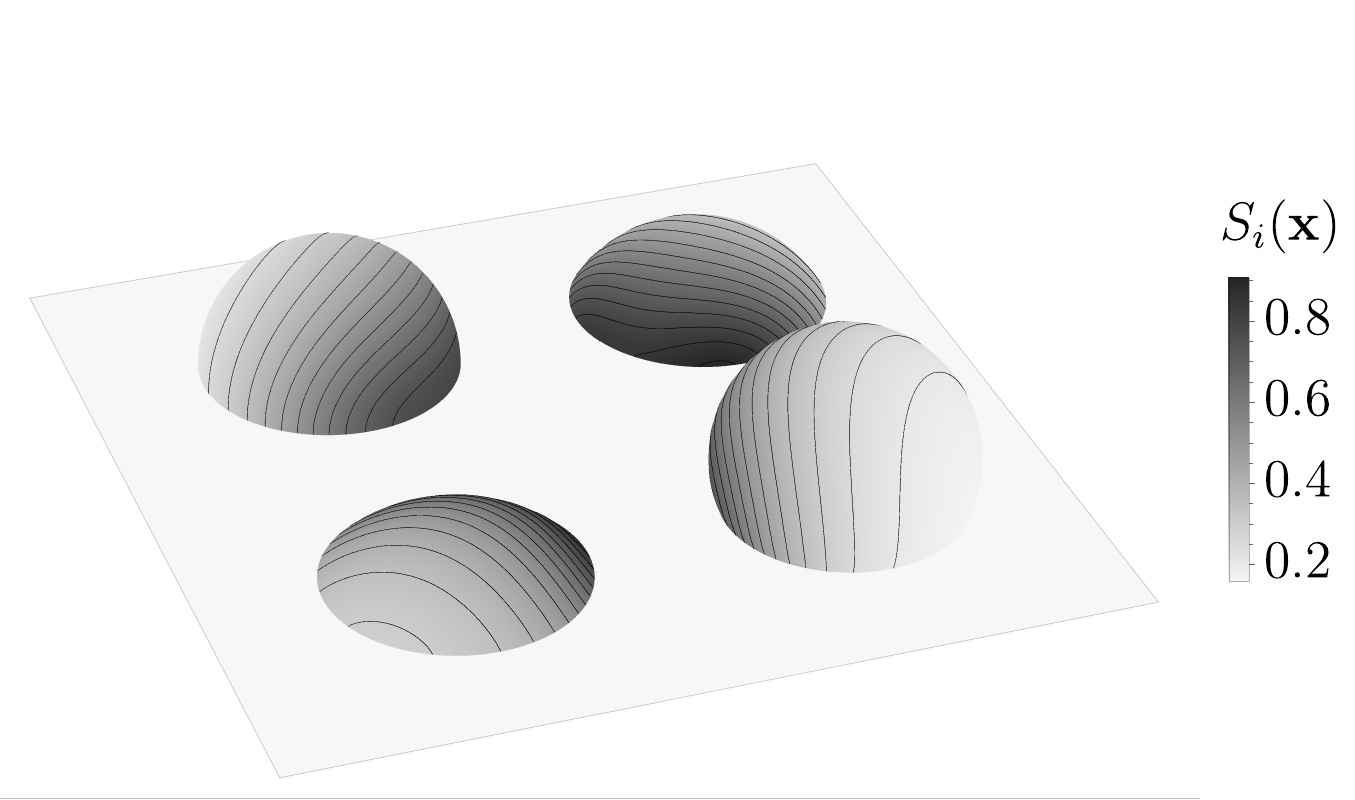} \\
(a) & (b) 
\end{tabular}
\caption{
        Shielding of four droplets as determined by (a) FEM and (b) the leading-order solution \eqref{eq:multidropFormula}. 
        Contact angles are $\pi/2$ (top left), $\pi/3$ (top right), $\pi/4$ (bottom left) and $5\pi/8$ (bottom right). 
        The footprint centres form a square of side length 3. 
        Contours are at intervals of $0.05$ in $S_i$.
    }\label{fig:fourDroplets}
\end{figure}

Figure \ref{fig:fourDroplets} compares the shielding of four droplets as predicted by FEM and the leading-order solution \eqref{eq:multidropFormula}.
Shielding is strongest towards the centre of the array and decreases nonlinearly outwards. 
The two calculations show similar shielding distributions across all four droplets.

\section{Liquid transport}\label{sec:transport}

We again use the scalings of \citet{wilson2023evaporation} and the single-source setup of Section \ref{sec:singleSource}.
The droplet is assumed to be pinned and surface tension causes it to retain its spherical-cap shape during evaporation.
We determine a measure of its net horizontal liquid transport, which influences deposit asymmetry.
By symmetry, the net horizontal transport is parallel to $\mathbf{e}_x$, and we denote its $x$-component by $U$. 
For an incompressible liquid with velocity $\mathbf{u}$ occupying a domain $V$, the divergence theorem gives
\begin{equation}
    U=\iiint_V \mathbf{u}\cdot\mathbf{e}_x \ud V=\iiint_V \nabla\cdot(x\mathbf{u}) \ud V=\iint_{\partial V} x\mathbf{u}\cdot\mathbf{n}\ud S.
\end{equation}
Using impermeability on the substrate, the kinematic condition in the form $ \mathbf{u}\cdot\mathbf{n}=\mathcal{V}_n+J^\mathrm{ps}$ on $\Gamma$, where $\mathcal{V}_n$ is the normal velocity of the interface, and axisymmetry of $\Gamma$, $\mathcal{V}_n$ and $J^{\mathrm{iso}}$,
\begin{equation}
    U=\iint_{\Gamma} x\left(\mathcal{V}_n+J^{\mathrm{ps}}\right)\ud {S}=\iint_{\Gamma} x\,J^{\mathrm{iso}}(\mathbf{x})[1-S(\mathbf{x})] \ud {S}=-q\iint_{\Gamma} x\frac{\rho^2}{\ell^3}J^{\mathrm{iso}}(\mathbf{x}^\ast)\ud S.
\end{equation}
Using the inversion \eqref{eq:inversion} and evaluating the standard single-layer potential,
\begin{equation}
    U=-q\iint_{\Gamma} J^{\mathrm{iso}}(\mathbf{x})\left( b+\rho^2\q{}{b} \right)\frac{1}{\ell} \ud S
    =-2\pi q \rho \dd{}{b}\left[ \rho C^{\mathrm{iso}}(b) \right]=-\rho q\dd{}{b}\left(\rho\frac{\Delta F}{q}\right),\label{eq:USol}
\end{equation}
where $\Delta F=2\pi q C^\mathrm{iso}(b)$ is the total shielding, equal to the reduction in the total evaporation rate due to the point source.
Hence $U$ is determined by the variation of the total shielding with source separation, geometrically weighted by $\rho$. 
In the far field, this scales as $b^{-2}$ for any contact angle.

\section{Conclusions}

We have derived a simple asymptotic theory for spatially resolved evaporation from interacting spherical-cap droplets using multipole expansions and Kelvin inversion. 
The exact response of the target droplet to a point source reveals a universal spatial structure, including a tangent ring separating more strongly and weakly shielded regions. 
Comparisons with finite-element calculations show errors decreasing with separation, with the largest errors occurring for the most narrowly separated dewetting cases.
Resolving the local evaporative flux also enables calculation of evaporation-driven liquid transport.
For a single source, we explicitly relate the total shielding to the net horizontal liquid transport.

\begin{bmhead}[Acknowledgements]
AWW thanks Stephen K. Wilson, Madeleine R. Moore, Hannah-May D'Ambrosio and Benjamin D. Goddard for insightful discussions and comments on this work.
\end{bmhead}

\begin{bmhead}[Declaration of interests]
The author reports no conflict of interest.
\end{bmhead}

\begin{bmhead}[Data availability statement]
The data that support the findings of this study are available from the corresponding author upon reasonable request.
\end{bmhead}

\bibliographystyle{jfm}
\bibliography{multiDropEvap}

\end{document}